# The Secular and Rotational Brightness Variations of Neptune

Richard W. Schmude Jr.

Gordon State College, 419 College Dr., Barnesville, GA 30204 USA

schmude@gordonstate.edu

Ronald E. Baker

Indian Hill Observatory, Chagrin Valley Astronomical Society,

PO Box 11, Chagrin Falls, OH 44022 USA

rbaker52@gmail.com

Jim Fox

Astronomical League and the AAVSO, P.O. Box 135, Mayhill, NM 88339 USA

makalii45@gmail.com

Bruce A. Krobusek

5950 King Hill Drive, Farmington, NY 14425 USA

bkrobusek@gmail.com

Hristo Pavlov

9 Chad Place, St. Clair, NSW 2759, Australia

hristo_dpavlov@yahoo.com

Anthony Mallama

14012 Lancaster Lane, Bowie, MD 20715 USA

anthony.mallama@gmail.com

2016 March 29

Abstract

Neptune has brightened by more than 10% during the past several decades. We report on the analysis of published Johnson-Cousins B and V magnitudes dating back to 1954 along with new U, B, V, R, $R_C$, I and $I_C$ photometry that we recorded during the past 24 years. Characteristic magnitudes, colors and albedos in all seven band-passes are derived from the ensemble of data. Additionally, 25 spectra spanning 26 hours of observation on 5 nights are analyzed. The spectrophotometry demonstrates that planetary flux and albedo is inversely related to the equivalent widths of methane bands. We attribute the changes in band strength, flux and albedo to the high altitude clouds which rotate across the planet's visible disk. Bright clouds increase albedo and flux while reducing methane absorption. Synthetic V magnitudes derived from the spectroscopy also agree closely with the photometric quantities, which cross-validates the two techniques. The spectroscopic and photometric results are discussed within the framework of the secular brightness variations of Neptune.

1. Introduction

This is the seventh paper in a series which models the brightness, color and albedo of all the planets except Earth on the Johnson-Cousins magnitude system. Mercury (Mallama et al. 2002) and Venus (Mallama et al. 2006) were observed with the Solar and Heliospheric Observatory satellite. The SOHO magnitudes augmented ground-based observations at illumination phases where the planets are too near the Sun to observe through the atmosphere. Mercury was found to have a very strong brightness surge when fully illuminated due to coherent backscattering from its airless surface. The phase curve of cloudy Venus has no brightness surge when fully lit but there is an anomalous peak around phase $170^o$ caused by forward scattering of sunlight by droplets of sulfuric acid. The results for all the other planets rely on ground-based observations which covered decades of time and a great variety of planetary aspects. Variations in the brightness of Mars (Mallama 2007) were found to depend on its rotational and orbital phase angles as well as the illumination phase. The brightness of Jupiter (Mallama and Schmude 2012) was fairly constant though cloud belt phenomena produced small changes. The inclination of the rings of Saturn (Schmude 2011 and Mallama 2012) dominates that planet's variations. Uranus (Schmude et al. 2015) is subject to large seasonal brightness changes especially at red and near-IR wavelengths. Methods for using photometric phase curves to characterize exo-planets have been outlined as well (Mallama, 2009).

Here we examine the photometric variations of Neptune, which has brightened over time for reasons that are still not clearly understood. The motivation for this study is to characterize Neptune in all seven of the Johnson-Cousins band-passes. Section 2 discusses the medium-band b- and y-filter observations of Lockwood and Jerzykiewicz (2006). In section 3 we describe our wide-band U, B, V, R, $R_C$, I and $I_C$ photometric methods. Section 4 gives the wide-band results including rates of secular brightness change along with reference magnitudes, colors and albedos. Our quantitative spectroscopic methods are described in section 5. Then, in section 6 we demonstrate that the equivalent widths of the methane bands are variable on a rotational time-scale and that the widths are inversely related to the planet's integrated flux and albedo. The spectroscopic and photometric results are also cross-validated there. The observations are discussed in the context of Neptune's atmosphere in section 7 and the paper is summarized in section 8.

## 2. Medium-band photometry at blue and yellow wavelengths

The main subject of this paper is wide-band photometry of Neptune in visible and near-IR wavelengths. However, the medium-band blue and yellow photometry reported by Lockwood and Jerzykiewicz (2006) anticipates some of our results. So, we review their work first.

Lockwood's web page at http://www2.lowell.edu/users/wes/U_N_lcurves.pdf updates the 2006 paper, providing data through 2014 and spanning 42 years. The annual means of their b (4720 Angstrom) and y (5510 Angstrom) magnitudes reveal an interval of brightening between about 1980 and 2000. The upward trends then taper off between 2000 and 2005 and subsequently the magnitudes remain nearly constant. The values after 2005 are about 0.12 magnitudes brighter than those of 1980.

The brightening happened between the times of Neptunian equinox and the occurrence of southern solstice. Sromovsky et al. (2003) had thus attributed the increase to a seasonal effect. However, Lockwood and Jerzykiewicz disputed that interpretation in their paper because earlier observations were not well represented by the seasonal model.

Hammel and Lockwood (2007) have suggested that the brightening may be related to an increase in solar irradiance. Nonetheless, they also mentioned that the Neptunian brightness increase is more than 100 times larger than the irradiance increase.

The variability of Neptune will be revisited in section 7. The data discussed above is plotted there in Figs. 7 and 8 along with our data and other photometry from the literature obtained at similar wavelengths.

## 3. Photometric method

The wide-band magnitudes examined in this study are on the Johnson-Cousins system. The original system established by Johnson et al. (1966) was expanded by Cousins (1976a and 1976b). The effective wavelengths and the full width at half maximum of the seven bands are listed in Table 1.

Table 1: Characteristics of the Johnson-Cousins systems

| Filter | Effective Wavelength (Angstroms) | FWHM (Angstroms) |
|---|---|---|
| U | 3600 | 680 |
| B | 4360 | 980 |
| V | 5490 | 860 |
| R | 7000 | 2090 |
| Rc | 6410 | 1580 |
| I | 9000 | 2210 |
| Ic | 7980 | 1540 |

In order to study the long-term brightness variations of Neptune, we retrieved measurements dating back to the 1950s reported by Serkowski (1961) and by Jerzykiewicz and Serkowski (1966). These data are referenced to a set of 'ten year standard' stars which have previously been shown to be consistent with the primary UBV standards and they are considered to be reliable. These data sources and the others described below are summarized in Table 2.

Schmude (1992, 1993, 1995, 1996, 1997a, 1997b, 1998a, 1998b, 2000a, 2000b, 2001, 2002, 2004, 2005, 2006a, 2006b, 2008, 2009, 2010a, 2010b, 2012, 2013, 2014, 2015a and 2015b) and co-workers report V filter results for Neptune. In many cases, B, R and I results are also listed. All observations were made differentially with respect to standard stars of the Johnson-Cousins system. The methodology is described in the papers listed above and it generally followed the procedures for recording data and for applying extinction corrections and color transformation described by Hall and Genet (1988). (Extinction and color transformation are described later in this section.) Some of the magnitudes reported in the earlier papers lacked color transformations but all the data analyzed here have been transformed. Formal errors are not available for these observations though they are estimated to be a few hundredths of a magnitude or less based on scatter in the data obtained over short periods of time as well as the likely uncertainties involved with extinction and transformation.

Photometry reported for the first time in this paper was acquired by two of the authors (BAK and REB) using 20- and 30-cm Schmit-Cassengrain telescopes, respectively. The observations were scheduled so that the air masses of the comparison star and of Neptune were nearly equal in order to minimize the correction needed for atmospheric extinction. The procedure for acquiring CCD image data for a single Johnson magnitude in a single filter was to record 2 separate series of 3 images of the planet interleaved with 2 such series for the standard star. The resulting magnitude for that filter represents the average of 6 values derived from 6 pairs of planet-and-standard CCD images. Thus, each set of 4 B, V, R, I (or B, V, $R_c$, $I_c$) magnitudes derives from 48 separate images. Additional flat field and dark frames were recorded at all observing sessions. Object and sky values were extracted using aperture photometry software, and those values were processed according to the methods outlined by Hardie (1962), including extinction correction and color transformation. The B, V, R and I magnitudes for the comparison star, 58 Aqr, were taken from Mendoza et al. (1966). The V-$R_c$ and V-$I_c$ colors (and the corresponding magnitudes) for 58 Aqr were derived from Mendoza's values by applying the transformations developed by Taylor (1986). Those colors were 0.18 and 0.36, respectively. The observations themselves are listed in Tables 3 and 4 along with their uncertainties.

Table 2: Data sets used in the development of the photometric model of Neptune

```
  Years        Filters       Source
1953-1961      B V           Serkowski (1961)
```

```
1962-1966      B V         Jerzykiewicz and Serkowski (1966)
1991-2015      B V R I     Schmude (1992, 1993, 1995, 1996, 1997a,
                           b, 1998a, b, 2000a, b, 2001, 2002, 2004,
                           2005, 2006a, b, 2008, 2009, 2010a, b,
                           2012, 2013, 2014 and 2015a, 2015b)
2014           B V R I     Krobusek (this paper)
2014           B V Rc Ic   Baker (this paper)
```

## Table 3: Magnitudes recorded by BAK

| MJD* | B | +/- | V | +/- | R | +/- | I | +/- |
|---|---|---|---|---|---|---|---|---|
| 56843.32 | -6.629 | 0.012 | -7.007 | 0.008 | -6.682 | 0.050 | -5.538 | 0.094 |
| 56885.26 | -6.630 | 0.009 | -6.999 | 0.015 | -6.688 | 0.048 | -5.532 | 0.092 |
| 56894.26 | -6.662 | 0.079 | -6.998 | 0.013 | -6.677 | 0.050 | -5.631 | 0.087 |
| 56895.24 | -6.601 | 0.011 | -7.018 | 0.009 | -6.691 | 0.050 | -5.486 | 0.096 |

```
*Modified Julian Date
 "+/-" refers to uncertainty
```

## Table 4: Magnitudes recorded by REB

| MJD* | B | +/- | V | +/- | Rc | +/- | Ic | +/- |
|---|---|---|---|---|---|---|---|---|
| 56870.30 | -6.613 | 0.018 | -7.003 | 0.011 | -6.772 | 0.018 | -5.872 | 0.043 |
| 56873.29 | -6.623 | 0.008 | -7.015 | 0.007 | -6.788 | 0.015 | -5.879 | 0.040 |
| 56873.30 | -6.629 | 0.008 | -6.993 | 0.013 | -6.793 | 0.015 | -5.900 | 0.039 |
| 56924.13 | -6.580 | 0.012 | -7.009 | 0.013 | -6.799 | 0.014 | -5.960 | 0.039 |
| 56924.14 | -6.593 | 0.006 | -7.008 | 0.008 | -6.778 | 0.016 | -5.959 | 0.037 |
| 56942.11 | -6.603 | 0.005 | -6.988 | 0.004 | -6.776 | 0.014 | -5.866 | 0.039 |
| 56942.12 | -6.623 | 0.004 | -7.003 | 0.005 | -6.790 | 0.014 | -5.872 | 0.040 |

```
*Modified Julian Date
"+/-" refers to uncertainty
```

As noted above, the magnitudes reported here were corrected for extinction and color effects. The extinction correction accounts for dimming due to the Earth's atmosphere by applying a correctional coefficient to the air mass through which the object was viewed. This results in 'natural' magnitudes. Color correction then converts the natural magnitudes to standard magnitudes on the Johnson-Cousins system. Equations 1 and 2 illustrate extinction correction and color transformation, respectively.

$$\Delta m_0 = \Delta m - k_M \, \Delta X$$

(1)

$$\Delta M = \Delta m_0 + \varepsilon \, \Delta C$$

(2)

In equation 1, $\Delta$ refers to the difference between Neptune and the standard star, $m$ is the raw magnitude, $k_M$ is the extinction coefficient, $X$ is the air mass and $m_0$ is the natural magnitude corrected for extinction. In equation 2, $\varepsilon$ is the transformation coefficient, $C$ is the color index and $M$ is the magnitude corrected for extinction and color.

A concern with regard to color transformation is that the color coefficients are derived from the observations of stars. However, the spectral profiles of stars are different from that of Neptune especially at long wavelengths where Neptune's profile is quite steep and is blanketed by methane bands as seen in Fig. 1.

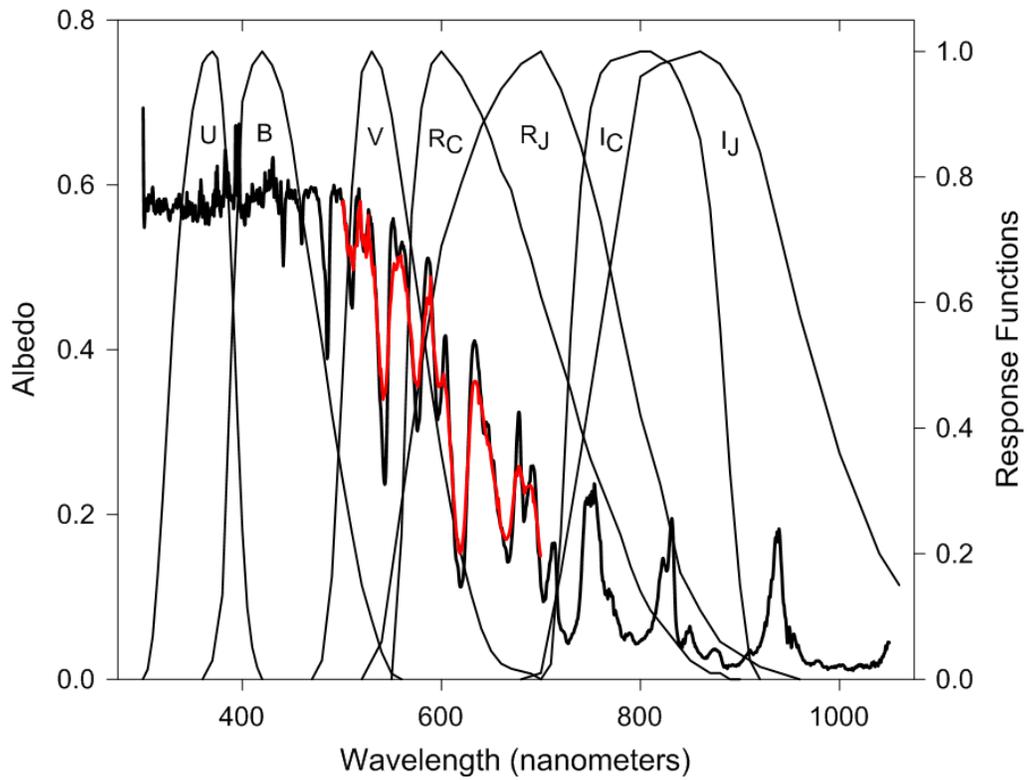

Fig. 1. The response functions of the Johnson-Cousins system (Filter Profile Service at http://skyservice.pha.jhu.edu) are superposed on the albedo of Neptune. The albedo shown in black from 300 to 1050 nm is from Karkoschka (1998). The albedo segment from 500 to 700 nm shown in red is from this paper.

After the reductions described above were applied, the magnitudes were corrected for distance. The absolute magnitudes reported in this study correspond to a distance of one astronomical unit from the Earth to Neptune and from the Sun to Neptune.

The effect of planetary oblateness has been corrected. Oblateness changes the apparent brightness of a body because its disc will have a greater area when observed (and illuminated) from the polar direction. Our magnitudes are for Neptune as observed from the equatorial direction.

The phase angle of Neptune can only be as large as two degrees, so its effect on the planet's magnitude is expected to be very small. Indeed, Karkoschka (2011) derived a phase angle coefficient of only 0.0028 magnitudes per degree from Hubble Space Telescope (HST) images though that figure excluded Neptunian clouds. In order to estimate the coefficient independently, we evaluated the phase functions of Jupiter in U, B, V, R and I (Mallama and Schmude, 2012) and of Venus in B, V, R and (Mallama et al., 2006) at the two degree limit for Neptune. In all cases the magnitude change is less than 0.01 magnitude. Likewise, Lockwood and Thompson (2009) found that the phase function for Titan is well under 0.01 magnitude per degree. These results are all consistent with results reported by Lockwood (1978) that the effect of solar phase angle is less than 0.003 magnitude in B. Since the effect of phase angle on the integrated brightness of Neptune is insignificant and not known precisely, no correction was applied. The mean magnitudes for each opposition are listed in Table 5.

Table 5: Mean magnitude and number of observations for each opposition

| Year | B | # | V | # | R | # | I | # | Rc | # | Ic | # |
|---|---|---|---|---|---|---|---|---|---|---|---|---|
| 1954.29 | -6.466 | 6 | -6.869 | 6 | ---- | 0 | ---- | 0 | ---- | 0 | ---- | 0 |
| 1955.30 | -6.469 | 4 | -6.861 | 4 | ---- | 0 | ---- | 0 | ---- | 0 | ---- | 0 |
| 1956.31 | -6.479 | 8 | -6.879 | 8 | ---- | 0 | ---- | 0 | ---- | 0 | ---- | 0 |
| 1957.31 | -6.482 | 8 | -6.882 | 8 | ---- | 0 | ---- | 0 | ---- | 0 | ---- | 0 |
| 1959.32 | -6.472 | 7 | -6.883 | 7 | ---- | 0 | ---- | 0 | ---- | 0 | ---- | 0 |
| 1960.33 | -6.477 | 10 | -6.913 | 10 | ---- | 0 | ---- | 0 | ---- | 0 | ---- | 0 |
| 1961.34 | -6.471 | 6 | -6.898 | 6 | ---- | 0 | ---- | 0 | ---- | 0 | ---- | 0 |

```
1962.34  -6.494  8   -6.918  8   ----   0   ----   0   ----   0   ----   0
1964.36  -6.482  7   -6.907  7   ----   0   ----   0   ----   0   ----   0
1965.36  -6.499 13   -6.899 13   ----   0   ----   0   ----   0   ----   0
1966.37  -6.498  8   -6.895  8   ----   0   ----   0   ----   0   ----   0
1991.52  -6.558 12   -6.920 32   ----   0   ----   0   ----   0   ----   0
1992.53  -6.524  8   -6.895 14   ----   0   ----   0   ----   0   ----   0
1993.53  -6.624 27   -6.906 28   -6.556 17   -5.407 17   ----   0   ----   0
1994.54  -6.622 11   -6.913 12   -6.561 11   -5.472 11   ----   0   ----   0
1995.55  -6.577  6   -6.965 10   -6.609  4   -5.331  5   ----   0   ----   0
1996.55  -6.591  5   -6.938 25   ----   0   ----   0   ----   0   ----   0
1997.56  ----   0   -6.952 20   ----   0   ----   0   ----   0   ----   0
1998.56  -6.597 11   -6.976 31   ----   0   ----   0   ----   0   ----   0
1999.57  -6.621 19   -6.993 39   -6.607  2   -5.607  2   ----   0   ----   0
2000.58  ----   0   -6.981  5   ----   0   ----   0   ----   0   ----   0
2001.58  -6.561  5   -7.003 57   ----   0   ----   0   ----   0   ----   0
2002.59  -6.570  3   -6.977 26   ----   0   ----   0   ----   0   ----   0
2003.60  -6.634 14   -6.990 16   ----   0   ----   0   ----   0   ----   0
2004.60  -6.577  3   -6.987  3   ----   0   ----   0   ----   0   ----   0
2005.61  -6.596 11   -6.995 12   ----   0   ----   0   ----   0   ----   0
2006.61  -6.594  5   -6.995  8   ----   0   ----   0   ----   0   ----   0
2007.62  -6.537 15   -6.971 16   ----   0   ----   0   ----   0   ----   0
2008.63  -6.600  8   -6.997 15   ----   0   ----   0   ----   0   ----   0
2010.64  -6.602 23   -7.000 24   ----   0   ----   0   ----   0   ----   0
2011.64  -6.629 18   -7.012 18   ----   0   ----   0   ----   0   ----   0
2012.65  -6.618 20   -7.001 21   ----   0   ----   0   ----   0   ----   0
2013.66  -6.610 21   -7.006 21   ----   0   ----   0   ----   0   ----   0
2014.66  -6.601 28   -6.995 33   -6.708  9   -5.651  6   -6.785  7   -5.900  7
```

## 4. Photometric results

This section begins with a discussion of the brightness increases measured in the B-, V-, R- and I-bands during the past several decades. Then reference values of magnitude, color and albedo are derived and reported.

The brightening of Neptune since 1954 is illustrated in Fig. 2 and the average rates of change are listed in the column labeled *Slope* in Table 6. The short wavelength B- and V-band observations extend over a greater time span than do the long wavelength R- and I-band data. Therefore the slopes of the short-wavelength magnitudes are evaluated over their entire duration (1954-2014) and over the same duration as the long wavelength data (1993-2014). This distinction is indicated in the column labeled *Time_Span.* The rates of brightening at long wavelength significantly exceed those of the short wavelengths over both time spans. The column labeled *Mag_1984* contains the magnitudes derived by evaluating the best fitting lines in the figure at the mid-time (year 1984 = MJD 45963) of the span for reasons that are made clear below.

Table 6. Magnitudes as a function of time

| Band | Time_Span | Mag_1984 | +/- | Slope* | +/- |
|------|-----------|----------|------|--------|------|
| B | 1954-2014 | -6.55 | 0.00 | -2.41 | 0.13 |
| V | " | -6.94 | 0.00 | -2.23 | 0.10 |
| R | 1993-2014 | -6.50 | 0.01 | -7.10 | 0.57 |
| I | " | -5.32 | 0.03 | -11.50 | 2.36 |
| B | 1993-2014 | -6.59 | 0.01 | -0.64 | 0.39 |
| V | " | -6.91 | 0.00 | -3.77 | 0.23 |

* Milli-magnitudes per year

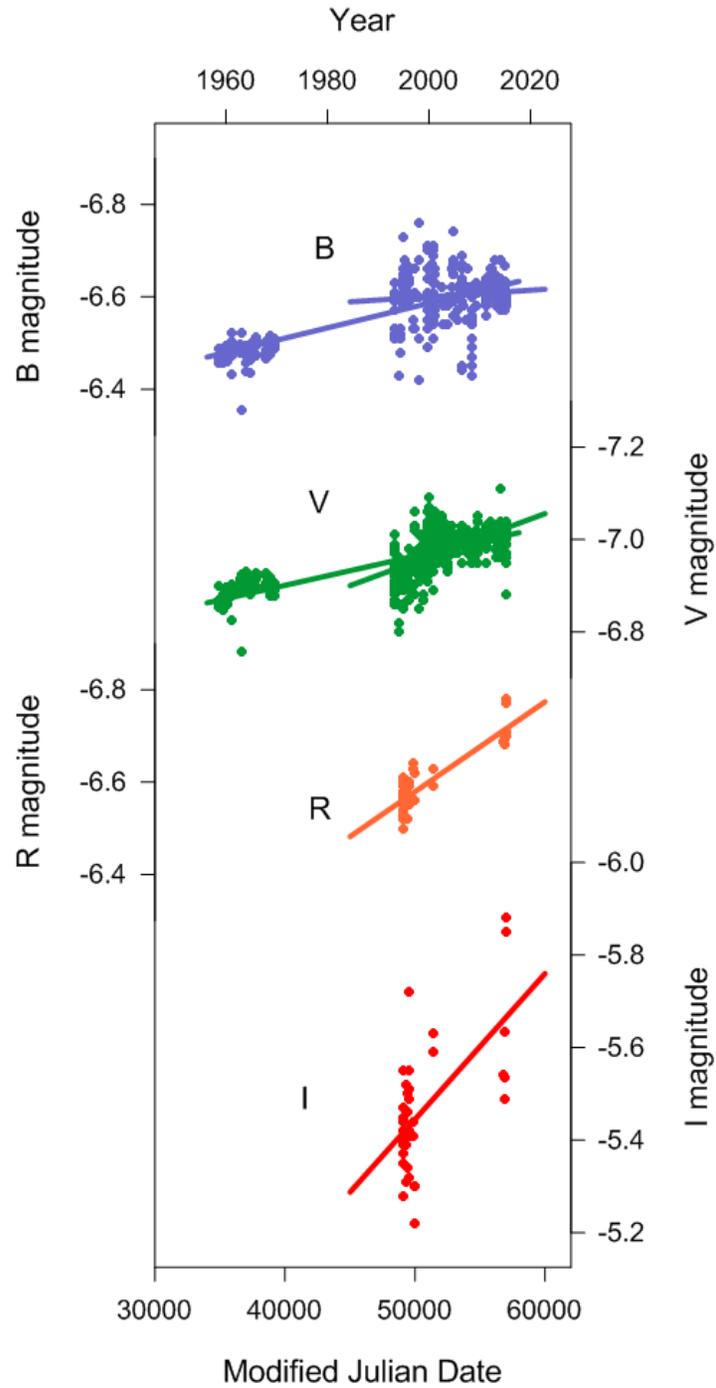

Fig. 2. Secular brightening of Neptune in the B-, V-, R- and I-bands. The short-term variations in the I band magnitudes are probably due to intermittent clouds crossing the visible disk of Neptune.

The determination of reference magnitudes for Neptune is important from an astronomical point of view. Reference works such as the Astronomical Almanac, Allen's Astrophysical Quantities and numerous astronomy text books list the magnitudes for the planets. In many cases these are out of date. Additionally, exo-planet research often refers to 'Neptunes' in a prototypical sense. So, it is worthwhile to characterize the brightness of the real Neptune.

The secular brightness increase of Neptune complicates the determination of reference magnitudes and colors. The adopted B and V magnitudes in Table 7 were derived by evaluating the best fitting lines in Fig. 2 for the period 1954-2014 at the mid-time of all the observations. We take these to be the most representative values based on the available data. Similarly, the R and I magnitudes were determined by projecting their best fitting lines to the same mid-time. The formal uncertainties for these bands listed in Table 6 have been increased in Table 7 to represent more realistic values. These take into account systematic effects including uncertainties in the magnitudes of comparison stars, in the color transformation and in the atmospheric extinction coefficients.

In addition to the long-term photometric data described above, one of the authors (REB) obtained $R_C$ and $I_C$ magnitudes in 2014. ($R_C$ and $I_C$ are characterized in Table 1 and the individual magnitudes are listed in Table 4.) The average values of those data from Table 5 are listed in Table 7 in the row labeled '2014'. The reference values of $R_C$ and $I_C$ listed in Table 7 were determined by applying the slopes from the closely related R and I data in Fig. 2 (listed in Table 6) and solving for the magnitudes at the mid-time. The uncertainties are realistic estimates (as described above) rather than formal errors.

Another of the authors (RWS) acquired U-band photometry in 2014. The average absolute magnitude during this interval was −6.46. The reference magnitude in Table 7 was derived by applying the slope from the B magnitudes in Fig. 2 and evaluating the U magnitude at the mid-time. The realistic uncertainty is estimated as above.

Table 7 also lists magnitudes from the Astronomical Almanac, and for the first and last years of observation (1954 and 2014) where they are available. The U, B and V Almanac magnitudes are all fainter than the reference magnitudes of 1984. However, the B and V magnitudes agree very closely with the observations of 1954 which suggests the data of Serkowski (1961) are the basis of the Almanac values. The brightening in the B and V magnitudes between 1954 and 2014 is about 0.14 magnitude.

Table 7. Reference magnitudes and related quantities

|                | U     | B     | V     | R     | Rc    | I     | Ic    |
|----------------|-------|-------|-------|-------|-------|-------|-------|
| Reference      | -6.38 | -6.55 | -6.94 | -6.50 | -6.61 | -5.32 | -5.71 |
| Uncertainty    | 0.05  | 0.02  | 0.02  | 0.05  | 0.05  | 0.08  | 0.08  |
| Astr. Almanac  | -6.25 | -6.46 | -6.87 | ---   | ---   | ---   | ---   |
| Difference     | -0.13 | -0.09 | -0.07 | ---   | ---   | ---   | ---   |
| 1954 (average) | ---   | -6.47 | -6.87 | ---   | ---   | ---   | ---   |
| 2014 (average) | -6.47 | -6.61 | -7.00 | -6.71 | -6.79 | -5.66 | -5.91 |

The geometric albedo is defined as the ratio of the observed flux for a planet to that of a perfectly reflecting Lambertian disk. Taking the V filter as an example, the magnitude of the Sun is −26.75 (Table 8), so the ratio of the luminosity of Neptune (magnitude −6.94 from Table 7) to that of the Sun is $1.19 \times 10^{-8}$ as shown in Equation 3.

$$\text{Lratio} = 10^{\{(-26.75+6.94)/2.5\}} = 1.19 \times 10^{-8}$$

(3)

The average disk radius of Neptune, *r*, is 24552 km and the astronomical unit distance is 149.6 × $10^6$ km. Hence there is an area factor, $\sin^2(r/AU)$ or $2.69 \times 10^{-8}$ and the geometric albedo, p, is 0.442 as shown in Equation 4.

$$p = Lratio/ \sin^2 (r/AU) = 0.442$$

(4)

Table 8. Solar magnitudes

| U* | B* | V* | R* | $R_C$** | I* | $I_C$** |
|---|---|---|---|---|---|---|
| -25.90 | -26.10 | -26.75 | -27.29 | -27.15 | -27.63 | -27.49 |

```
 * Livingston (2002)
** Binney and Merrifield (1998).
```

The albedo values listed in Table 9 range from a high of 0.575 in the U-band to a low of 0.046 in the I-band. This skewing is also reflected in the colors of Neptune which are described in the next paragraph. The albedos of Uranus are discussed in section 7.

Table 9. Geometric albedos

| | U | B | V | R | Rc | I | Ic |
|---|---|---|---|---|---|---|---|
| Reference | 0.575 | 0.562 | 0.442 | 0.181 | 0.225 | 0.046 | 0.072 |
| Uncertainty | 0.029 | 0.011 | 0.008 | 0.004 | 0.011 | 0.001 | 0.004 |
| 1954 | --- | 0.522 | 0.415 | --- | --- | --- | --- |
| 2014 | 0.625 | 0.594 | 0.467 | 0.218 | 0.266 | 0.060 | 0.086 |
| Uranus | 0.500 | 0.561 | 0.488 | 0.202 | 0.263 | 0.053 | 0.089 |

The color indices listed in Table 10 reveal the peculiar energy distribution of Neptune. The planet is moderately red according to the U-B and B-V indices but quite blue according to V-R and R-I. The presence of strong methane bands accounts for these unusual colors. The B-V index changed by only 0.01 magnitude between 1954 and 2014.

Table 10. Colors

|  | U-B | B-V | V-R | V-Rc | R-I | Rc-Ic |
|---|---|---|---|---|---|---|
| Reference | +0.17 | +0.39 | -0.44 | -0.33 | -1.18 | -0.90 |
| Uncertainty | 0.05 | 0.03 | 0.03 | 0.05 | 0.04 | 0.07 |
| 1954 | --- | +0.40 | --- | --- | --- | --- |
| 2014 | +0.14 | +0.39 | -0.29 | -0.21 | -1.05 | -0.88 |
| Uranus | +0.32 | +0.50 | -0.42 | -0.27 | -1.12 | -0.84 |

5. Spectrophotometric method

The spectroscopic data for this study were recorded by author HP with the 35-cm telescope at Tangra Observatory (IAU E24) using a TACOS video system (Gault, D. et al., 2014). A Star Analyzer grating of 100 lines per mm produced a dispersion of 12 Angstroms per pixel on the WAT-910BD analogue video camera which was operated with gamma set to 1.0 and gain set to 32 dB.

Videos containing between 300 and 1000 frames were recorded using the OccuRec video recording software (Pavlov 2015a) running in stacking mode. In this mode the signal from an accumulating video camera, in this case the WAT-910BD, is recorded as images with up to 14 significant bits. The recording software provided specialized features to ensure that the spectra are not saturated and that the zero order image was placed in the same area of the chip for all videos. Video exposures from 0.160 to 2.56 s were used, depending on the brightness of the object, to ensure maximum signal-to-noise ratio while avoiding saturation.

During each night of observation spectra of several standard stars from the CalSpec library (Bohlin et al. 2014) were recorded along with those of Neptune. The spectral energy distributions (SEDs) of the standards are derived from data obtained with the Space Telescope Imaging Spectrograph on HST. The absolute accuracy of fluxes in the CalSpec library is estimated to be about 1% based on comparisons between synthetic magnitude derived from the spectra and the Johnson-Cousins magnitudes of photometric standard stars (Bohlin and Landolt, 2015).

Pairs of instrumental fluxes and calibrated wavelengths were extracted from the video images using the Tangra software package (Pavlov 2015b). Images were corrected for dark frame background and then the individual frames were aligned within a sub-pixel precision centered on the fitted PSF of the zero order image. A third-order wavelength calibration was performed for one of the stars displaying well defined Balmer absorption lines using the zero order image, the H$\delta$, H$\gamma$, H$\beta$ and H$\alpha$ lines and the two telluric $O_2$ lines at 6869 and 7605 Angstroms. The remaining stars and Neptune were then wavelength calibrated based on the position of their zero-order image and the previously determined calibration polynomial. The accuracy of the wavelength calibration is estimated to be about four Angstroms.

The wavelength-calibrated instrumental fluxes where then exported to the AbsFlux program (Mallama et al. 2015) which performed the absolute radiometric calibrations. AbsFlux simultaneously models instrument sensitivity and atmospheric extinction, as functions of wavelength, based on least-squares fitting between instrumental fluxes and the corresponding CalSpec library fluxes. The standard stars were observed over a wide range of elevations so that the software could separate the effect of extinction from that of sensitivity. The parameters of the model where then applied to the instrumental fluxes of Neptune in order to generate absolute SEDs. Two examples are shown in Fig. 3.

Comparisons between observed fluxes and CalSpec library fluxes for the same stars indicate that the radiometric uncertainty of the data is about 5% RMS on a per-wavelength basis (Mallama et al., 2015) though it is smaller when averaged over a span of wavelengths. The uncertainties include contributions from modeling errors, Poisson noise and a small non-linearity of the detector response.

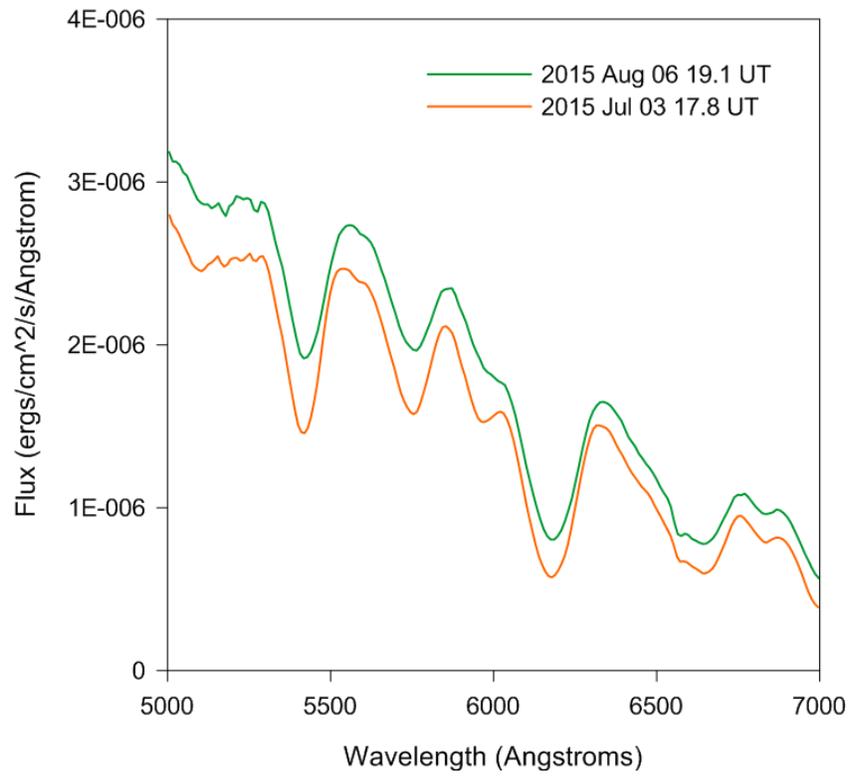

Fig. 3. The brightest and faintest SEDs recorded during 26 hours of observing on 5 nights. The variations are discussed in section 6.

## 6. Spectrophotometric results

The absolute SEDs of Neptune from 5000 to 7000 Angstroms were analyzed in detail because there are significant methane absorption bands present and because that interval is within the range of the peak sensitivity of the WAT-910BD video camera which extends from 4000 to 7000 Angstroms. Both the sensitivity of the video camera and the flux of Neptune decline at longer wavelengths which results in low signal-to-noise.

In order to validate our spectrophotometric fluxes we took an average over the 25 spectra by sorting them into 200 bins of 10 nm width. The binned averages were then converted to albedos and compared with the values reported by Karkoscha (1998). The solar spectrum used in the conversion (retrieved from http://rredc.nrel.gov/solar/spectra/am0/) is based on the work of Wehrli (1985) and Neckel and Labs (1981). Analysis of the results of the comparison, illustrated in Fig. 4, indicate an offset of -0.002 (in the sense our data minus Karkoschka's) and an RMS difference of 0.034. The very small offset is somewhat surprising given that Karkoscka's data was recorded in 1995 and that Neptune has brightened in the intervening years by about 0.04 or 0.07 magnitude based on the slopes from Table 6. Compensating factors that might be involved include Neptune's diurnal variability and the difference in the spectroscopy standard stars employed. Karkoschka quotes a 4% absolute accuracy for the albedos. In any case, we consider the small offset to be rather fortuitous but generally supportive of accurate spectroscopy. The larger RMS value is due mostly to the smaller bin sizes (and higher spectral resolution) of the albedos reported in Karkoschka's data set. That is, our data does not capture all of the depths of the spectral minima nor all of the heights of the maxima. The average albedos over the 5000 to 7000 range are listed in Table 11. Note that these do not correspond directly with the V magnitudes in that table which are discussed next. These albedos are evenly weighted over the interval specified above while the V magnitudes are weighted according to the response function shown in Fig. 1.

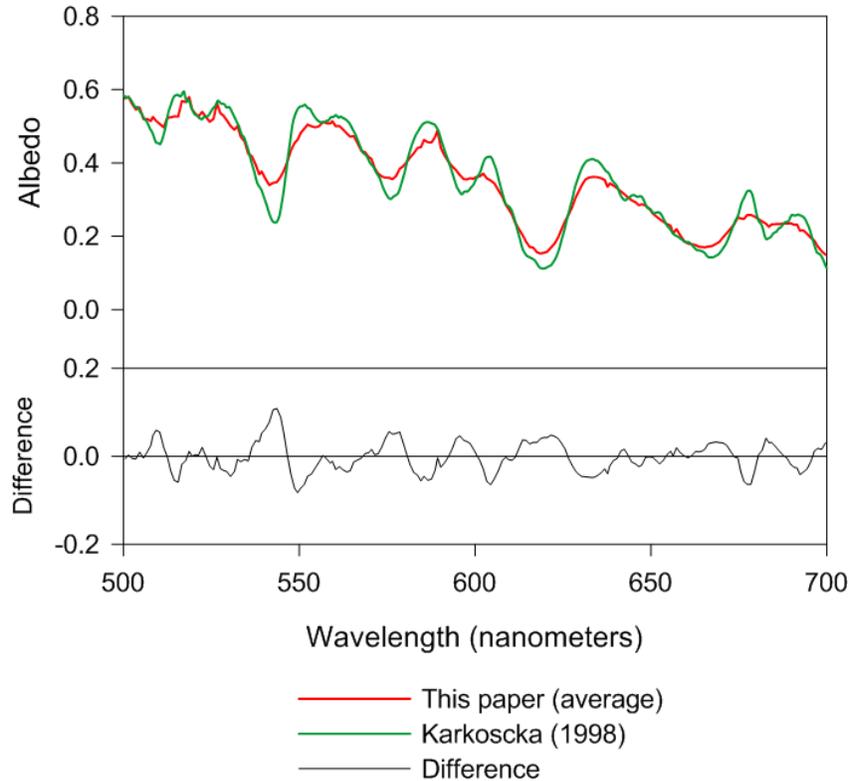

Fig. 4. The average of the spectrally resolved albedos from this paper is compared with that of Karkoschka (1998). The differences between them are shown in the bottom panel.

As a second method of validation, we derived synthetic V magnitudes from our fluxes. Since the sensitivity of the V band extends from 4700 to 7000 Angstroms (see Johns Hopkins Filter Profile Service at http://skyservice.pha.jhu.edu) we had to include data short of the 5000 Angstrom limit discussed above. The fluxes were convolved with the V filter sensitivity following the integral equation given by Smith et al. (2002) and Fukugita et al. (1996). The resulting V magnitudes are listed in Table 11. The average of those values, -6.97, is consistent with the average of our photometric V magnitudes for 2014, -7.00, listed in Table 7. Year-to-year and diurnal variations may account for the 0.03 magnitude difference.

Having validated the spectrophotometric fluxes and the resulting albedos, we performed the following geophysical analysis. The equivalent widths of four strong bands and the integrated fluxes over that entire wavelength range were measured. Band centers and the corresponding pairs of continuum wavelengths used to measure their equivalent widths were 5425 (5275 and 5525), 5760 (5650 and 5850), 6190 (6050 and 6350), and 6680 (6550 and 6750) Angstroms. The integrated fluxes and equivalent width values are listed in Table 11.

Then, the fluxes and the albedos from Table 11 were plotted as a functions of the summed equivalent widths as illustrated in Fig. 5. The best fitting lines have negative slopes that are significant at more than three standard deviations. We discuss the implications of this finding in the next section.

Table 11. Equivalent widths, fluxes, synthetic magnitudes and albedos

```
------ UT ------    -- Equivalent Widths --   --- Flux  ---   V-Mag   Albedo
Year Mon Da Hour    5425A 5760A 6190A 6680A   (ergs/cm^2/s)   -----   ------
2015 Jul 03 15.2    45.1  29.3  90.5  35.4      3.274E-03      -6.96   0.343
2015 Jul 03 17.8    47.9  29.1  91.1  37.4      3.078E-03      -6.89   0.322
2015 Jul 18 14.2    38.0  26.7  91.7  36.9      3.259E-03      -6.94   0.343
2015 Jul 18 15.5    34.1  23.2  89.7  35.2      3.252E-03      -6.93   0.343
2015 Aug 06 12.1    39.3  26.0  89.3  34.7      3.282E-03      -6.95   0.344
2015 Aug 06 13.0    40.3  28.1  91.4  36.9      3.318E-03      -6.96   0.348
2015 Aug 06 14.1    38.8  27.7  93.8  38.3      3.372E-03      -6.97   0.354
2015 Aug 06 15.0    37.7  26.8  92.7  36.3      3.365E-03      -6.96   0.354
2015 Aug 06 16.0    33.4  24.5  93.2  37.3      3.462E-03      -6.98   0.364
2015 Aug 06 17.1    35.5  24.1  86.6  33.6      3.578E-03      -7.03   0.376
2015 Aug 06 18.0    34.6  20.9  80.4  29.7      3.589E-03      -7.04   0.376
2015 Aug 06 19.1    34.7  21.0  78.4  28.3      3.597E-03      -7.05   0.377
2015 Aug 10 11.6    35.2  21.3  79.6  28.0      3.493E-03      -7.03   0.367
2015 Aug 10 12.6    37.9  23.7  82.1  30.2      3.413E-03      -7.00   0.358
2015 Aug 10 13.5    35.5  24.0  87.7  33.5      3.307E-03      -6.96   0.348
2015 Aug 10 16.8    31.5  17.7  75.6  27.0      3.383E-03      -6.98   0.356
2015 Aug 10 18.6    33.1  22.9  84.6  31.4      3.377E-03      -6.98   0.355
2015 Aug 13 11.5    37.5  23.2  90.1  35.7      3.307E-03      -6.94   0.348
2015 Aug 13 12.5    35.8  22.5  92.4  37.6      3.301E-03      -6.93   0.348
2015 Aug 13 13.5    39.9  26.2  92.4  39.1      3.347E-03      -6.95   0.352
2015 Aug 13 14.5    41.6  27.4  93.5  39.3      3.490E-03      -7.00   0.367
2015 Aug 13 15.6    34.7  24.1  93.4  37.4      3.340E-03      -6.93   0.352
2015 Aug 13 16.4    35.7  24.6  93.5  37.7      3.369E-03      -6.94   0.355
2015 Aug 13 17.5    35.6  24.7  91.6  38.3      3.340E-03      -6.93   0.352
2015 Aug 13 19.0    38.4  24.6  85.9  37.6      3.349E-03      -6.96   0.352
```

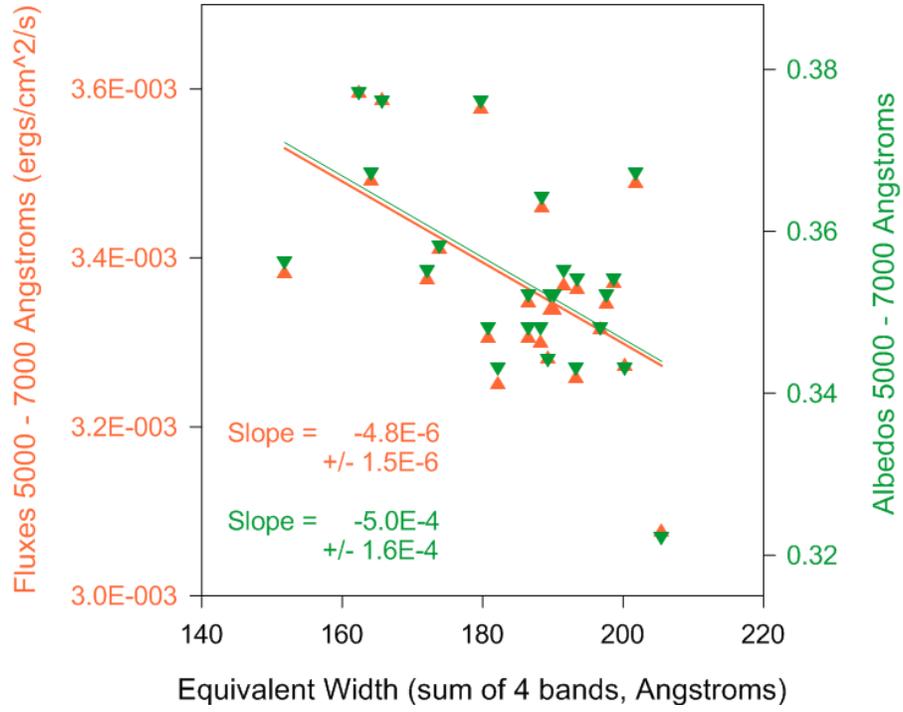

Fig. 5. Integrated flux and albedo as functions of equivalent width.

7. Discussion

This section begins by highlighting a few aspects Neptune's atmosphere which pertain to our study, especially the bright clouds and aerosols. Then we place our findings into the context of what is known and what is unknown about the atmosphere.

The bright clouds were characterized by Hammel et al. (1995) while Rages et al. (2002) described a bright apparition of Neptune's south polar feature. Sromovsky et al. (1993), Hammel and Lockwood (1997), and others have investigated cloud dynamics. Gibbard et al. (2003) reported that the altitudes of these clouds place them in the stratosphere and upper troposphere. Both Sromovsky et al. (2001) and Karkoschka (2011) have remarked that a cloud may change very rapidly, even over a few hours. Thus, a feature seen at a particular longitude may be difficult to identify after just one 16 hour rotation of the planet. Karkoschka's conclusion was based on a comprehensive study of HST images.

Some recent papers have reported high-resolution observations obtained with adaptive optics. A study combining Keck near- and mid-IR data with VLA radio observations by de Pater et al. (2014) led those authors to conclude that there are two distinct cloud layers. The deep layer is located in the troposphere at pressure levels of a few hundred mbar while a higher 'spatially intermittent' layer lies in the stratosphere at 20-30 mbar. Irwin et al. (2014) discuss 'image cubes' of 64x64 pixels and 2048 wavelengths in the J- and H-bands obtained with ESO's Very Large Telescope in Chile. In both 2009 and 2013 they recorded bright clouds at mid-latitudes, lower-elevation transient clouds near the equator and small discrete clouds around the pole.

Luszcz-Cook et al. (2014) discuss the many unknown and poorly understood characteristics of the Neptunian atmosphere. These include the general question of variability and structure of tropospheric and stratospheric aerosols, the relationship of discrete clouds to the aerosols, the number and optical thicknesses (and possible variability) of the cloud decks below the one bar level and whether hydrogen sulfide is a major constituent, the possible existence of a vertically extended haze, an optically thin methane haze at 1-2 bar and photochemical hydrocarbons that may contribute to one or more thin stratospheric layers of haze.

The presence of bright clouds, possibly at more than one altitude, is a recurrent theme in the literature synopsized above. Our spectroscopic results are consistent with discrete clouds at high elevations. The finding that brightness is an inverse function of equivalent width can be explained by the high albedo clouds reflecting sunlight before it is absorbed by the methane below. Similarly, Flagg et al. (2015) have demonstrated that material from the impact of comet SL-9 on Jupiter in 1994 reflected IR light that would otherwise have been absorbed by methane.

Direct evidence for the effect of clouds on brightness also comes from disk-resolved images of Neptune recorded with the Wide Field Planetary Camera (WFPC) on HST. Fig. 6 combines images recorded on 2011 June 25 with a graph of the planet's disk-integrated brightness. A variation of 0.08 magnitude occurs as a result of high-albedo clouds moving across the disk over a period of 12 hours. The planetary brightness reaches maximum when the clouds are near its meridian. Simon et al. (2015) found similar results from Hubble 8450 Angstrom data but with about twice the amplitude. This may indicate more and brighter clouds in recent years.

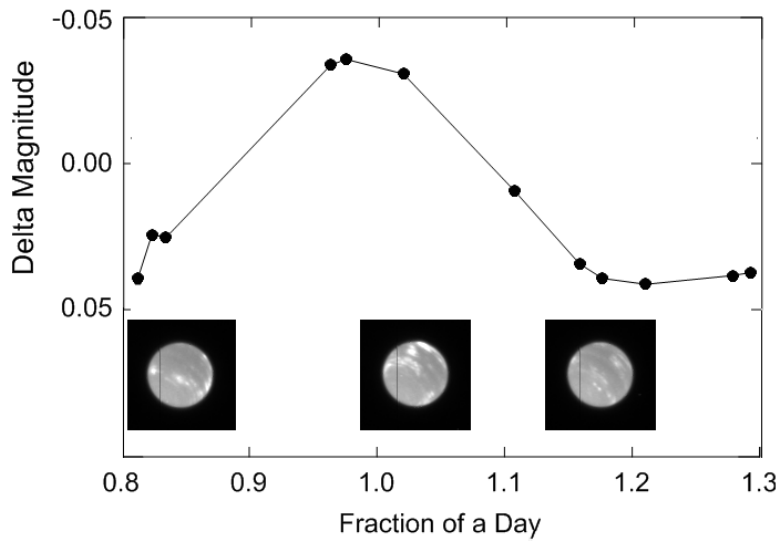

Fig. 6. HST/WFPC images of Neptune recorded at 8450 Angstroms reveal bright clouds crossing the disk while aperture photometry indicates a corresponding brightness variation. The images are associated with proposal number 12675 in the HST MAST archive.

Moving on from rotational brightness variations to secular changes, Karkoschka (2011) suggested that the brightening of Neptune is due to settling of dark aerosols to lower levels in the atmosphere. Furthermore, he postulated occasional 'darkening events' where dynamical events elevate haze particles to higher altitudes.

The secular variation is quantified by the b- and y-band photometry of Lockwood and Jerzykiewicz (2006), B and V photometry from the literature analyzed in this paper and the new multi-band observations that we have recorded over the past two decades. The combined short-wavelength results (B with b, V with y), which span 36 per cent of Neptune's 165 year orbital period, are shown in Figs. 7 and 8, along with magnitudes from Karkoscka (2011).

The observations suggest that brightening has leveled off in recent years. This might indicate that dark haze particles have settled completely. Additionally, the albedo data in Table 9 indicates that the combined reflectivity of Neptune in the B- and V-bands is now roughly equal to that of Uranus (which has a similar spectrum) while in 1954 it was substantially less than that planet. This result, too, may indicate that  Neptune is now in its un-darkened state. In any case, continued photometric monitoring might provide the initial alert that a major new darkening event (the first to be observed) is underway. These disk-integrated observations would also be important for characterizing the dimming phenomenon itself at wavelengths from blue through the near-IR.

Additionally, it seems that clouds might play a role in the secular variation in addition to their demonstrable rotational effect. Karkoschka (2011) found a significant variation of discrete cloud cover on a time scale of about 5 years from HST data obtained from 1994 through 2008. The albedos shown in his Fig.3 also appear to be influenced by Neptune's clouds. Disk-resolved imaging is the most direct method of detecting clouds. However, if equivalent widths are a proxy for cloudiness, as suggested by our spectroscopy, then such observations may also play a role in understanding the darkening phenomenon.

Finally, it is worth noting that increasing numbers of amateur astronomers are capable of imaging the Neptunian clouds. The Association of Lunar and Planetary Observers (http://alpo-

 ) hosts an archive of such images. These could be correlated with observations of variability in disk-averaged photometry and even with equivalent-widths.

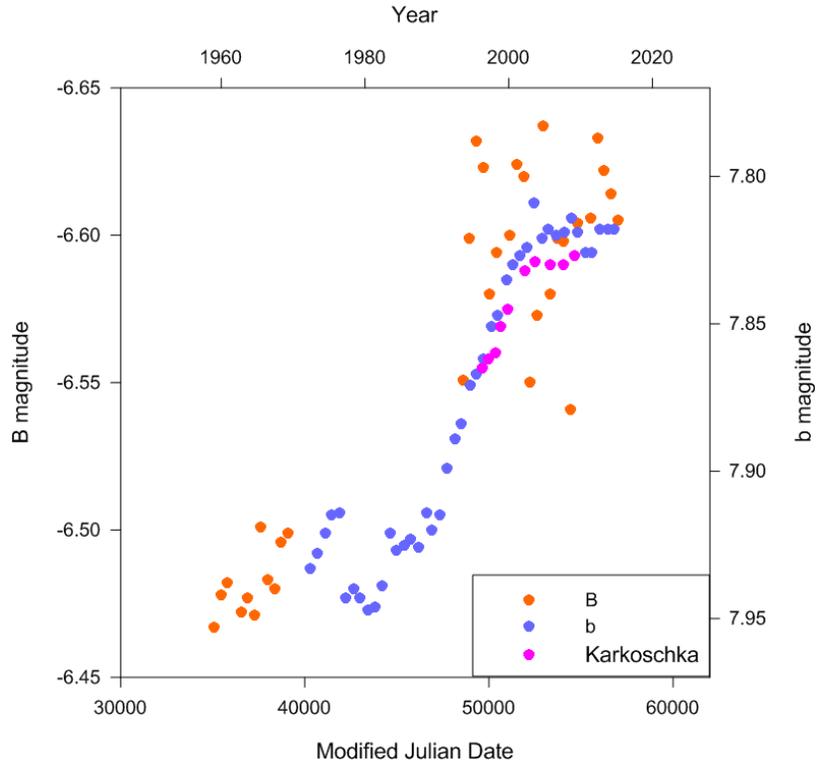

Fig. 7. Blue magnitudes. The two Y axes have been placed such that the average B- and b-band magnitudes in the brightness plateau region between 2005 and 2014 are aligned. B values are at a distance of one AU while b values are at the mean opposition distance of Neptune. The B data are from this paper and from the general literature (see section 3). The b magnitudes are from Lockwood (see section 2). The values from Karkoscka (2011) were digitized from his Fig. 2 and plotted on the b magnitude axis.

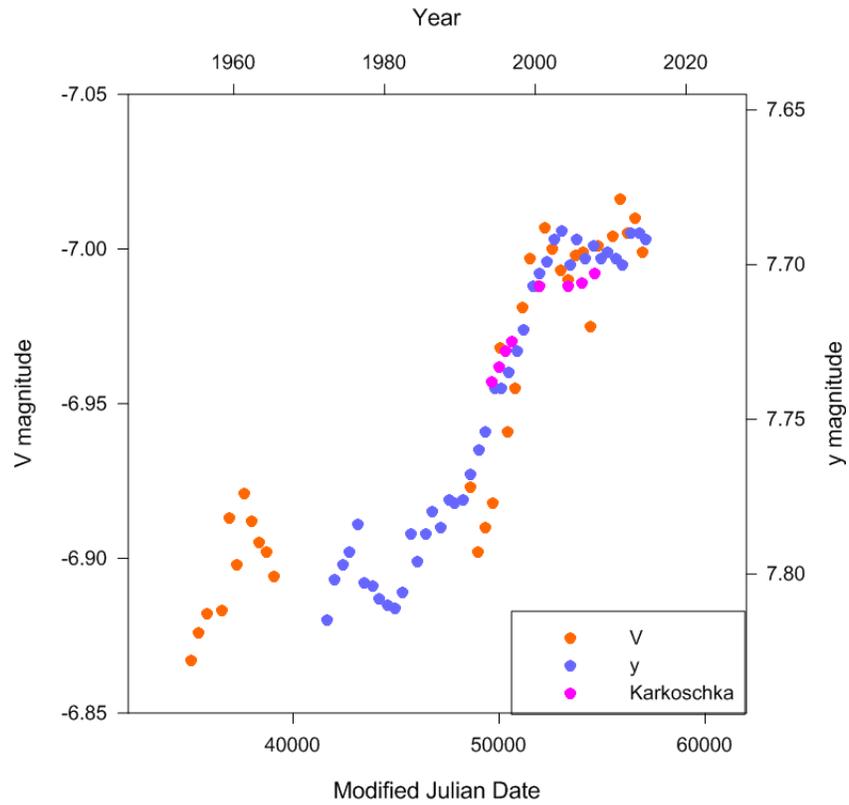

Fig. 8. Same as Fig. 7 but for yellow/green magnitudes, that is, the V- and y-bands. The V data are from this paper and from the literature (see section 3). The y magnitudes are from Lockwood (see section 2). The values from Karkoscka (2011) were digitized from his Fig. 2 and plotted on the y magnitude axis.

8. Summary

New photometric and spectrophotometric data are presented. The results of the two techniques have been cross-validated. The observations of Neptune span near-UV through near-IR wavelengths. These data are combined with B- and V-band magnitudes from the literature to illustrate a secular brightening of about 0.14 magnitude. Reference magnitudes, colors and albedos in all seven bands of the Johnson-Cousins system are also derived from the photometry.

The b- and y-band photometry of Lockwood and Jerzykiewicz (2006) indicates that most of Neptune's brightness increase occurred in the 20 years beginning around 1980. Our longer wavelength R- and I-band photometry demonstrates a higher rate of brightening since the observations began in 1993 as compared to our short-wavelength data. Continued photometric monitoring could reveal a future 'darkening event' of the kind hypothesized by Karkoschka (2011).

Spectroscopic observations demonstrate an inverse relationship between the integrated flux (and the albedo) over the 5000-7000 Angstrom range as a function of equivalent widths of methane absorption features. This result is consistent with high-altitude clouds that increase the overall planetary albedo and simultaneously reduced the effect of methane absorption.